# Spin Seebeck effect and ballistic transport of quasi-acoustic magnons in room-temperature yttrium iron garnet films

Timo Noack[1], Halyna Yu. Musiienko-Shmarova[1],
Thomas Langner[1], Frank Heussner[1], Viktor Lauer[1],
Björn Heinz[1], Dmytro A. Bozhko[1], Vitaliy I. Vasyuchka[1],
Anna Pomyalov[2], Victor S. L'vov[2], Burkard Hillebrands[1]
and Alexander A. Serga[1]

[1] Fachbereich Physik and Forschungszentrum OPTIMAS, Technische Universität Kaiserslautern, 67663 Kaiserslautern, Germany
[2] Department of Chemical Physics, Weizmann Institute of Science, Rehovot 76100, Israel

E-mail: `tnoack@rhrk.uni-kl.de`

**Abstract.** We studied the transient behavior of the spin current generated by the longitudinal spin Seebeck effect (LSSE) in a set of platinum-coated yttrium iron garnet (YIG) films of different thicknesses. The LSSE was induced by means of pulsed microwave heating of the Pt layer and the spin currents were measured electrically using the inverse spin Hall effect in the same layer. We demonstrate that the time evolution of the LSSE is determined by the evolution of the thermal gradient triggering the flux of thermal magnons in the vicinity of the YIG/Pt interface. These magnons move ballistically within the YIG film with a constant group velocity, while their number decays exponentially within an effective propagation length. The ballistic flight of the magnons with energies above 20 K is a result of their almost linear dispersion law, similar to that of acoustic phonons. By fitting the time-dependent LSSE signal for different film thicknesses varying by almost an order of magnitude, we found that the effective propagation length is practically independent of the YIG film thickness. We consider this fact as strong support of a ballistic transport scenario – the ballistic propagation of quasi-acoustic magnons in room temperature YIG.

## 1. Introduction

A permanently growing interest in the field of spin-caloritronics, which combines thermoelectrics with spintronics and nanomagnetism, underlines the importance of spin currents [1] as an alternative to charge currents for the utilization in logic devices [2, 3, 4]. This is due to zero Joule heating and the wide spectrum of methods to generate and manipulate spin currents. The spin current may arise in the form of charge currents with opposite flow directions for spin up and spin down carriers, or it can consist of magnons – the quanta of collective spin excitations [3, 5].

Among other methods, the magnon current can be created by a thermal gradient induced in a ferromagnet exposed to a magnetic field [6, 7]. This effect is referred to as the longitudinal spin Seebeck effect (LSSE) [8, 9]. Despite extensive studies, the exact microscopic mechanism responsible for the magnon-mediated spin Seebeck effect (SSE) is not yet completely clarified. In particular, some fundamental transport properties of magnetic materials, related to the LSSE, require further experimental and theoretical clarification. Among them is the timescale of the formation of the LSSE. The temporal dynamics of the LSSE is tightly connected with fundamental properties of a magnon gas such as the magnon mean free path. This physical quantity is crucial for the understanding of the transport properties of magnetic materials and the general peculiarities of magnon-phonon interaction [10, 11, 12, 13, 14, 15], as well as for the engineering of efficient LSSE-based spin-caloritronic devices [16, 17].

Typically, the spin Seebeck effect is indirectly detected by an electric voltage in a thin film of a heavy normal metal (e.g. Pt, Pd, W) deposited on the surface of a magnetic material. This voltage appears as a result of the conversion of a spin current into an electric one by the inverse spin Hall effect (ISHE) due to spin-dependent electron scattering. The temporal characteristics of ISHE voltages are determined both by the electron dynamics [18] in the metal and by the magnon dynamics in the magnetic media.

It has been experimentally demonstrated [19] by using coherently excited magnons, that in low-damping magnetic materials, such as epitaxial YIG films, the temporal profile of the ISHE voltage is dominated by the magnon dynamics in the magnetic insulator, rather than by the very fast electron dynamics in the normal metal. Similarly, it has been shown [20] that the LSSE dynamics is strongly influenced by the transport of thermal magnons inside the magnetic material and, thus, depends on the temporal development of the temperature gradient in the magnetic material close to the YIG/Pt interface.

Recently, a number of studies on the time dependency of the LSSE in YIG/Pt bilayers have been pre-sented [20, 21, 22, 23, 24, 25] and a rise time of the LSSE-induced ISHE voltage ($V_{\mathrm{LSSE}}$) of a few hundred nanoseconds was reported. Furthermore, a magnon propagation length of about $L \sim 500\,\mathrm{nm}$ was determined from the time-dependent measurements [20, 22] in a $6.7\,\mu\mathrm{m}$ thick YIG film, using a diffusion model of thermally driven magnons. However, these values of the propagation length are not unanimous and vary in the literature [20, 23, 25, 26, 27, 28] between several hundred nanometers and a few micrometers.

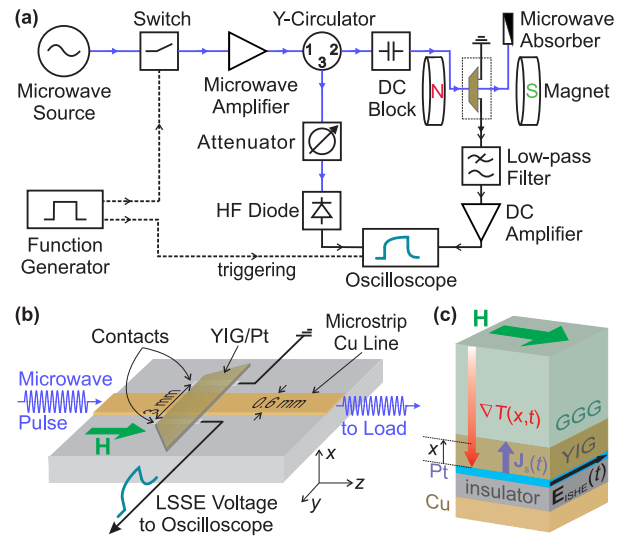

**Figure 1.** a) A schematic view of the experimental setup used for the investigation of the temporal evolution of the LSSE voltage. b) Sketch of the sample holder with the mounted YIG/Pt sample. c) Structure of the YIG/Pt sample, showing the relative orientation of the bias magnetic field **H**, the thermal gradient $\nabla T(t,x)$ created by the microwave heating of the platinum cover, the magnon carried spin current $\mathbf{J}_s(t,x)$, and the electric field $\mathbf{E}_{\mathrm{ISHE}}(t)$ induced in the Pt layer due to the inverse spin Hall effect [9].

In this article, we study the transient evolution of the LSSE voltage $V_{\mathrm{LSSE}}$ in YIG films of different thicknesses. Using a modified magnon transport model [20] with the time decay defined by a ballistic flight model, we find that the magnon propagation length is practically independent from the YIG film thickness and represents a material property of YIG.

## 2. Experimental setup

A schematic representation of the experimental setup is shown in Fig. 1. In this experiment, the microwave-induced heating of the Pt layer is used to create the thermal gradient in YIG/Pt bilayers [22]. The Pt film is heated by eddy currents, induced in the metal by a microwave field. This field is created by a $600\,\mu\mathrm{m}$ wide microstrip transmission line placed below the samples. In all measurements, the microstrip is driven by a microwave generator that produces microwaves at a fixed frequency of 6.875 GHz.

The 10 µs-long microwave heating pulses are applied with a repetition rate of 1 kHz to provide the system with sufficient time for cooling down after every heating cycle. The applied microwave power is set to 30 dBm. To avoid possible reflections of microwave energy, a matched 50 Ohm load is connected at the end of the microstrip line. The samples are placed on top of the microwave antenna with the platinum layer facing downwards [see Fig. 1(b)]. A few micrometer thick insulating layer was used to prevent galvanic contact between the platinum and the microstrip line.

All samples used in the experiments have the same structure, shown in Fig. 1(c): The YIG films of different thicknesses were grown in the (111) crystallographic plane by liquid phase epitaxy on a 500 µm thick gadolinium gallium garnet (GGG) substrate. A 10 nm-thick Pt layer was deposited on top of the YIG films using sputter deposition.

The thermal gradient generates a spin current across the YIG/Pt interface, shown in Fig. 1(c) by a violet arrow. Due to the spin-dependent scattering of spin polarized conducting electrons in the Pt layer, this spin current is converted into a charge current in-plane to the metal layer. This effect is known as the inverse spin Hall effect (ISHE) and leads to an electric potential perpendicular to the external magnetic field. Therefore, the resulting DC-voltage is proportional to the number of magnons transferring its angular momentum to electrons at the YIG/Pt interface. The ISHE voltage is given by: $V_{\text{ISHE}} \propto \Theta_{\text{SHE}} \left( \mathbf{J}_{\text{S}} \times \sigma \right) l$, where $\Theta_{\text{SHE}}$ is the spin Hall angle, which defines the efficiency of the ISHE in platinum, $\mathbf{J}_{\text{S}}$ is the spin current, $\sigma$ is the spin polarization and $l = 3$ mm is the distance between the electric contacts, as shown in Fig. 1(b).

Due to the small amplitude of the ISHE voltage of several microvolts, the electric signal from the YIG/Pt sample is amplified by using a DC voltage amplifier. Before the amplification, the signal is sent through a low-pass filter (DC-400 MHz) to avoid a possible disturbance of the sensitive receiving circuit by strong microwave pulses. Finally, the pulsed DC signal is displayed on the oscilloscope together with a reference microwave pulse, which has been reflected from the sample and afterwards directed to a HF diode detector by a microwave circulator (cf. Fig. 1(a)).

Figure 1(b) shows the geometry of the sample holder. The magnetic field **H** is oriented in the sample plane along the microstrip direction. The magnetic field strength is $H = 250$ Oe for all measurements. This value was chosen to avoid both resonant and parametric excitations of spin waves in the YIG film by the microwave magnetic field. The absence of such excitation processes is evidenced by the fact that the shapes of the observed LSSE pulses were identical for 250 Oe and for over 2500 Oe, where the spin-wave spectrum is shifted up so high that the frequency of the input microwave pulses lies well below the bottom of the spectrum.

## 3. Time-resolved investigation of the YIG thickness dependent temporal behavior of the longitudinal spin Seebeck effect

The main goal of this work is the investigation of the time dependent behavior of the LSSE voltage for different YIG thicknesses. The experiment has been carried out for eleven different samples with YIG thicknesses between 150 nm and 53 µm. For each sample, the evolution of the LSSE-induced ISHE voltage $V_{\text{LSSE}}$ resulting from a 10 µs-long heating pulse is analyzed. In order to eliminate nonmagnetic thermoelectric contributions to $V_{\text{LSSE}}$, the LSSE-induced ISHE voltage was measured twice for opposite orientations of the magnetic field $H$. Next, the difference $V_{\text{LSSE}}^{+\text{H}}$ - $V_{\text{LSSE}}^{-\text{H}}$ is calculated. Since the LSSE changes its sign with the magnetic field orientation, this voltage difference gives twice the value of the LSSE voltage with accompanying static effects be removed.

The normalized time profiles of LSSE voltages for the selected YIG thicknesses are shown in Fig. 2. The time evolution of the voltages clearly depends on the YIG-layer thickness. For the thinnest samples with a YIG thickness of 150 nm (not shown) and 300 nm, almost rectangular profiles are observed (cf. Fig. 2). For these thicknesses, the voltage is rising almost instantaneously and reaches its saturation level already after a few nanoseconds. With increasing YIG thickness, the voltage profiles deviate from the rectangular shape. The slower rising time and later saturation of the sig-nals is clearly correlated with the YIG film thickness in the range from 150 nm to 23 µm. In contrast, for the three thickest YIG films (23 µm, 30 µm and 53 µm) we observe practically identical LSSE signal profiles. These results can be understood by considering the motion of the magnons in the sample. The LSSE is a magnon transport process driven by the temperature gradient, leading to a spatial distribution of the thermally excited magnons. A flow of magnons arises to compensate this spatial inhomogeneity. Since the heating of the Pt-layer is homogeneous in the contact area, the thermal gradient is created perpendicular to the interface. Thus, the diffusion process is also oriented perpendicular to the interface and therefore limited by the thickness of the YIG layer. The magnons, excited farther away from the YIG/Pt interface, have to propagate a different distance than the magnons excited in its vicinity. Hence, they contribute to the LSSE voltage with smaller amount at a later point in time. This effect is particularly pronounced for thick YIG samples. Additional influence on the transient behavior of the LSSE voltage can be attributed to the thickness-dependence of the evolution of the thermal gradients in the samples. The contributions of these factors are discussed below.

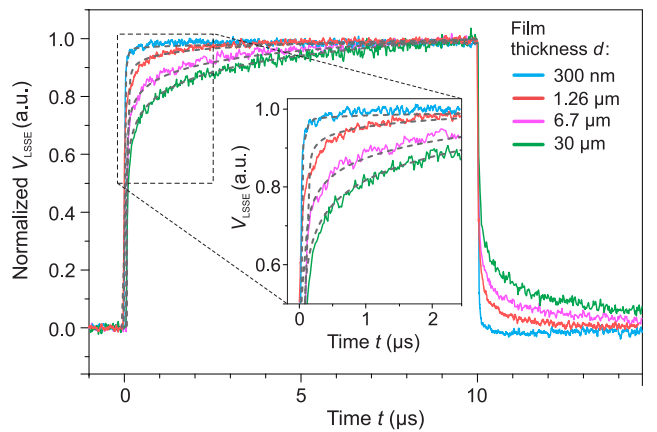

**Figure 2.** Temporal evolution of $V_{\text{LSSE}}$ for different YIG thicknesses $d$. Four dependencies $V_{\text{LSSE}}(t, d)$ for $d = 0.3, 1.26, 6.7$ and $30$ µm are shown by solid lines. The LSSE voltage build-up becomes visibly slower with increasing film thickness. The dashed lines represent the simulation of the LSSE dynamics using Eq. 1.

## 4. Extraction of the effective magnon propagation length

To obtain more quantitative predictions regarding the $V_{\text{LSSE}}$ time dependence and the magnon propagation lengths, we performed simulations of the heat dynamics for the sample structure shown in Fig. 1(c). This was done by using the numerical simulation software COMSOL Multiphysics® and by solving the one-dimensional heat transport problem analytically. The used thermal parameters of the materials are taken from Ref. [25]. The results of both calculations are nearly identical and consistent with the previous calculations in Ref. [20]: The rather slow temperature changes in the YIG film, developing on the millisecond time scale, are accompanied by a fast nano- and microsecond dynamics of the temperature gradient $\nabla T$, which strongly depends on the distance from the YIG/Pt interface $x$ (see Fig. 3).

In Refs. [20, 25], the spin Seebeck voltage was considered as a combination of interface and YIG bulk effects. Taking into account that the rise time of the thermal gradient at the interface is much faster than the observed rise time of the LSSE and as a first attempt to understand the data, we use here only the bulk effect for the fitting of the measured data:

$$V_{\text{LSSE}}(t) \propto \int_0^d \nabla T_x(x,t) \exp\left(\frac{-x}{L}\right) dx. \quad (1)$$

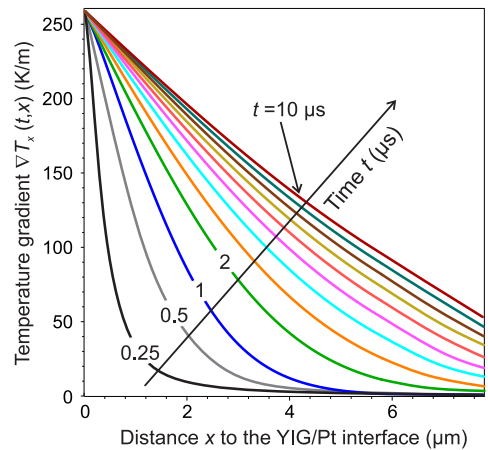

**Figure 3.** Spatial distribution of the temperature gradient $\nabla T_x(x,t)$ analytically calculated at different moments of time $t$ for the sample structure shown in Fig. 1(c).

Here $L$ is the effective magnon propagation length, $t$ is the time and $d$ is the thickness of the YIG film.

The length $L$ in Eq. 1 should be determined by some interactions in the YIG films. Without going into details of the magnon-magnon, magnon-phonon and other interactions, one expects that $L$ should be independent of the film thickness, as long as $L \ll d$. This expectation is a general physical requirement. For instance, in real gases, the mean free path of the atoms or molecules does not depend on the size of the container as long as it is much smaller than the container size. This requirement will serve us as a criterion in comparing different dynamical models for the LSSE signals evolution.

Using the simulated temporal evolution of $\nabla T$, we calculated the bulk term for different magnon propagation lengths $L$. Finally we used the method of least squares to determine the best value of $L$ for every profile. The measured dynamics of the normalized spin Seebeck voltage is plotted in Fig. 2 together with the simulated bulk terms dynamics. By using this fitting procedure, we determined the effective magnon propagation length $L$ for every YIG film thickness with an accuracy of 20 nm. The obtained values of $L$ are shown in Fig. 4 by empty blue diamonds. These values clearly increase with the thickness of the investigated YIG films.

The observed behavior obviously contradicts the previously mentioned physical requirement. Most likely it is caused by the fact that such an important factor as the magnon propagation dynamics is not accounted in Eq. 1. In fact, it is supposed that the magnons which are driven somewhere in the sample by a local temperature gradient are immediately contributing to the ISHE. In order to improve the model we need to involve the propagation time of these magnons to the YIG/Pt interface $\tau_{\text{m}}$ and to take into account their movement through a spatially varying temperature gradient. Following Hioki *et al.* [25], the magnon delay can be accounted by a simple modification of Eq. 1:

$$V_{\text{LSSE}}(t) \propto \int_0^d \nabla T_x(x, t - \tau_\text{m}) \exp\left(\frac{-x}{L}\right) dx. \qquad (2)$$

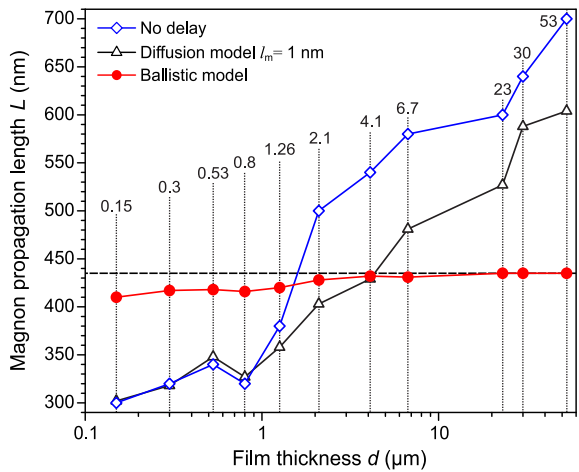

**Figure 4.** The magnon propagation length $L$, defined using three different models: The simple "no-delay" model [20, 22] Eq. (1) (empty blue diamonds), the magnon delay model Eq. (2) with diffusive magnon delay (empty black triangles) and the ballistic propagation model (filled red circles). The horizontal dashed line marks the saturation level $L = 435$ nm. The numbers near the vertical dotted lines represent the thickness of the measured YIG films in micrometers.

The exact form of $\tau_\text{m}$ was defined in Ref. [25] in the framework of a diffusion propagation model as $\tau_\text{m} = x^2/(l_\text{m} v_\text{m})$, where $l_\text{m}$ is the mean free path of the thermal magnons and $v_\text{m}$ is their group velocity. This velocity can be calculated using experimental [29, 30] and theoretical [31, 32, 33, 34, 35] data of the magnon spectra in YIG. It is known, that starting from about 1 THz ($\approx 20$ K) and up to the end of the first Brillouin zone at about 6.5 THz ($\approx 300$ K), the lowest magnon branch, which is mostly populated at room temperature, has an almost linear dispersion relation $\omega(\mathbf{k})$. Therefore, for the thermal magnons in this branch, the magnon velocity is constant in a wide frequency range. Its value along the [111] crystallographic direction, normal to the surfaces of all of our films, is about $v_\text{m} = 10^4$ m/s.

The question about the value of the magnon mean free path $l_\text{m}$ is much more complicated. In Ref. [15, 25], a value of $l_\text{m} \approx 1$ nm was assumed at room temperature. Taking these values of $l_\text{m}$ and $v_\text{m}$, we fitted our experimental data by Eq. 2. The resulting behavior of $L$, shown by empty black triangles in Fig. 4, is qualitatively the same as in the "no delay" case of Eq. 1. This result is not very surprising: The 1 nm value for $l_\text{m}$ was obtained by Boona and Heremans [15] by comparison of phonon and magnon contributions to the specific heat and to the thermal conductivity of a bulk YIG sample and by a consequent interpolation of the low temperature (2 K–20 K) magnon-related data to the room temperature range. However, the relevance of such an interpolation is not completely clear due to the strong differences in the relaxation and spectral characteristics of low- and high-energy magnons [36]. Boona and Heremans emphasized that their "estimate is conservative, especially at room temperature, where SSE experiments are typically conducted". We also have to notice that the diffusive delay in Eq. 2 accounts exclusively for the diffusive spreading of the magnon package. This makes the applicability of the simplified diffusion model in Eq. 2 rather questionable.

On the other hand, the linear dispersion relation of the lowest magnon mode allows us to suggest an alternative "ballistic" model of the magnon propagation. In the ballistic approach, the magnon delay is defined as $\tau_\text{m} = x/v_\text{m}$. At room temperature the magnons populate the lowest magnon mode and at high frequencies their velocity obeys the relation: $\omega(\mathbf{k}) \approx v_\text{m} k$. Indeed, for the "acoustic" magnons with $\omega(\mathbf{k}) = v_\text{m} k$, the conservation laws for the dominant four-magnon scattering

$$\omega(\mathbf{k}_1) + \omega(\mathbf{k}_2) = \omega(\mathbf{k}_3) + \omega(\mathbf{k}_4),$$
$$\mathbf{k}_1 + \mathbf{k}_2 = \mathbf{k}_3 + \mathbf{k}_4, \qquad (3)$$

are satisfied only if $\mathbf{k}_1 \parallel \mathbf{k}_2 \parallel \mathbf{k}_3 \parallel \mathbf{k}_4$, i.e. when all magnons propagate in the same direction, with the same velocity $\mathbf{v}_\text{m} = \mathbf{k}_1/k_1$. The same condition $\mathbf{k}_1 \parallel \mathbf{k}_2 \parallel \mathbf{k}_3$ is correct for the three-magnon processes

$$\omega(\mathbf{k}_1) + \omega(\mathbf{k}_2) = \omega(\mathbf{k}_3),$$
$$\mathbf{k}_1 + \mathbf{k}_2 = \mathbf{k}_3, \qquad (4)$$

as well as for the processes with any other number of magnons (see, e.g., Chapter 1 in the book [37]).

This means that the package of magnons propagates ballistically with the velocity $\mathbf{v}_\text{m}$ and all types of interaction processes within the "acoustic" magnon mode with the linear dispersion law do not change the direction and the value of the propagation velocity $\mathbf{v}_\text{m}$, leading only to an evolution of the package shape in the $\mathbf{k}$-space during the ballistic flight. In addition, the dominating four-magnon processes (Eq. 3) preserve the total number of magnons in the package, while the three-magnon processes, that change their number, are much less probable for the exchange magnons. In such a situation, the two-particle scattering on crystal defects alongside with the Cherenkov radiation [10] and the four-magnon interaction between the "acoustic" magnons and the "optical" magnons with $\omega(\mathbf{k}) \approx$ const. can be seen as the dominant mechanism, restricting the propagation length of the thermal magnons.

In view of the aforesaid, we used the ballistic model to fit the experimental $V_{\text{LSSE}}$ waveforms and to determine the propagation length $L$ for the different YIG thicknesses. The result is presented in Fig. 4 by the filled red circles. Due to the rather weak dependence of the obtained values on the YIG-film thickness, the length $L$ was determined with an accuracy of 1 nm. As it is clearly seen in the figure, as the YIG thickness grows by a fraction of eight (from 150 nm to 1.2 µm), the propagation length increases only by about 3.5% (from 410 nm to 428 nm). The magnon propagation length $L$ is therefore almost independent of the YIG film thickness and can be considered as an inherent property of YIG. Under such an assumption the propagation length should become independent of the film thickness for $L \ll d$, as observed: For the films with a thickness $d \geq 4.1\,\mu m$, $L$ saturates near 435 nm (see Fig. 4). This value agrees well with both our previous estimations [20, 22] and the results of other groups [26, 38]. The slight increase in $L$ in the smaller thickness range can be related to, e.g., the relative decrease in the density of crystal defects at large distances from the YIG/GGG interface in thicker epitaxial YIG films.

## 5. Summary

In this article, we studied the influence of the magnetic insulator thickness in YIG/Pt bilayers on the temporal dynamics of the longitudinal spin Seebeck effect (LSSE). A microwave-induced heating technique has been used to generate a thermal gradient across the bilayer interface. The experiment demonstrates a strong dependence of the time evolution of the LSSE signal on the magnetic layer thickness. An increase of the YIG thickness from 150 nm to 53 µm leads to a 7-fold increase in the rise time of the detected LSSE voltage. The experimental data have been precisely fitted using a model which assumes ballistic motion of thermal magnons in a temperature gradient.

The average magnon propagation length of about 425 nm was found to be almost independent of the YIG film thickness. This fact strongly supports the suggested simple ballistic model of the magnon propagation in room-temperature YIG films.

## Acknowledgments

Financial support by Deutsche Forschungsgemeinschaft (DFG) within Priority Program 1538 "Spin Caloric Transport" (project SE 1771/4-2) and DFG project INST 248/178-1 as well as technical support from the Nano Structuring Center, TU Kaiserslautern are gratefully acknowledged.